\documentclass[preprint,showpacs,preprintnumbers,amsmath,amssymb]{revtex4}
\usepackage{graphics}
\usepackage{dcolumn}
\usepackage{bm}

\begin{document}

\title{A unit vector for characterizing the spin polarization of free electron}

\author{Chun-Fang Li\footnote{Email address: cfli@shu.edu.cn} and Yan Wang}

\affiliation{Department of Physics, Shanghai University, 99 Shangda Road, Baoshan
District, 200444 Shanghai, China}


\begin{abstract}

New degrees of freedom having the form of a unit vector are identified for characterizing
the spin polarization of free electron. It is shown that when only the spin is
considered, the non-commutativity of the Cartesian components of the Pauli vector allows
us to use the azimuthal angle of a second direction, denoted by unit vector $\mathbf I$,
with respect to the quantization direction to characterize the spin polarization. The
rotation of $\mathbf I$ through an angle about the quantization axis leads to a rotation
of the spin polarization vector through twice the angle about the same axis. Discussions
are also made in Heisenberg picture as well. Upon utilizing this approach to a free
electron and letting the quantization direction for each plane wave be the wave vector,
we arrive at a representation in which the unit vector $\mathbf I$ functions as an
independent index to characterize the spin polarization.

\end{abstract}

\pacs{75.25.-j, 03.65.Fd, 02.20.Qs}
\maketitle


\section{Introduction}

In the past decade, the spin Hall effect of electron, a phenomenon that is usually
described as a transverse spin current \cite{Hirsch, Zhang, Kato, Valenzuela} generated
by an electric current, drew much attention of the spintronics community. But there is
still a controversy \cite{Rashba, Shi, Sonin, Tokatly} over the definition of spin
current. Since the spin Hall effect is closely related to the spin polarization, we are
concerned with the description the spin polarization of electron in continuous states.
The purpose of this paper is to advance a representation formalism for the spin
polarization of free electron.

To this end, we will first isolate the spin from other degrees of freedom and identify a
unit vector to characterize the spin polarization. As we know, when only the spin is
considered, its polarization is totally determined by a normalized spinor. The spin
polarization vector (SPV) $\mathbf s$ that corresponds to an arbitrary normalized spinor
$\chi$ is defined as \cite{Merzbacher}
\begin{equation} \label{definition of SPV}
\mathbf{s} = \chi^{\dag} \boldsymbol{\sigma} \chi,
\end{equation}
where $\boldsymbol {\sigma}$ is the Pauli vector. In fact, the SPV and its corresponding
spinor satisfy the following eigen equation,
\begin{equation} \label{property of SPV}
\mathbf{s} \cdot \boldsymbol{\sigma} \chi= \chi.
\end{equation}
In addition, the non-commutativity of the Cartesian components of the Pauli vector
requires a direction to determine the spatial quantization of the spin. This direction is
known as the quantization axis; and any state spinor $\chi$ can be expressed as a linear
superposition of the two mutually orthogonal eigen spinors with respect to this axis.
Apart from the quantization axis, we identify a second direction to characterize the SPV.
This direction is denoted by unit vector $\mathbf I$. It is shown that upon changing
$\mathbf I$ by a SO(3) rotation through an angle about the quantization axis, one
realizes a SU(2) rotation of the state spinor through twice the angle about the same
axis. As a result, the corresponding SPV is changed by a SO(3) rotation also through
twice the angle about the same axis. When applying this approach to a free electron by
means of the plane-wave expansion and assuming the quantization axis for each plane wave
to be the wave vector \cite{Beenakker}, we get a representation in which the unit vector
$\mathbf I$ plays its unique role in characterizing the spin polarization. To the best of
our knowledge, this is the first time to report on such a representation.

This paper is arranged as follows. An algebraic approach is advanced in Section
\ref{description of eigen spinors} to find the two mutually orthogonal eigen spinors of
spin operator $\mathbf{w} \cdot \boldsymbol{\sigma}$, where unit vector $\mathbf{w}$
denotes the quantization axis. From this approach we find in both the two eigen spinors
the same degree of freedom that takes on the form of a unit vector $\mathbf I$. It
describes the eigen spinors in such a way that a rotation of $\mathbf{I}$ about the
$\mathbf{w}$ axis results in opposite changes in their phases. As a result, any spinor
that is expressed as a superposition of the two eigen spinors is characterized by such a
degree of freedom. It is shown in Section \ref{description of SPV} that rotating unit
vector $\mathbf{I}$ through an angle will lead to a rotation of the SPV through
\textit{twice} the angle. The same result is obtained in Heisenberg picture as well. The
previous approach is applied in Section \ref{application} to obtain a representation of
free electron's spin polarization with the help of plane-wave expansion. Upon assuming
the quantization axis for each plane wave to be the wave vector, we arrive at a
representation, the two polarization bases of which are characterized by the same unit
vector as well as the same weighting function. The unit vector in this case is shown to
play an independent role in describing the spin polarization. Section \ref{conclusions}
concludes the paper with remarks.

\section{\label{description of eigen spinors} Characterization of the phase of
eigen spinors}

In this section, we will find a unit vector $\mathbf I$ to determine the phases
\cite{Zeng} of both the eigen spinors $\chi^w_{\pm}$ that satisfy the following equation,
\begin{equation} \label{basic spinors}
\mathbf{w} \cdot \boldsymbol{\sigma} \chi^w_{\pm} =\pm \chi^w_{\pm},
\end{equation}
with $\mathbf w$ being the quantization axis. In order to do so, let us first introduce
this vector to express $\chi^w_{\pm}$.

\subsection{An algebraic approach to obtain $\chi^w_{\pm}$: Introduction of unit vector $\mathbf I$}

Given a quantization axis $\mathbf w$, we are free to introduce a unit vector
$\mathbf{I}$ that together with $\mathbf{w}$ axis defines two unit vectors $\mathbf{u}$
and $\mathbf{v}$ in the following way,
\begin{subequations} \label{u and v}
\begin{align}
  \mathbf{u}& =\mathbf{v} \times \mathbf{w}, \\
  \mathbf{v}& =\frac{\mathbf{w} \times \mathbf{I}}{|\mathbf{w} \times \mathbf{I}|} \label{v}.
\end{align}
\end{subequations}
They form a Cartesian system with $\mathbf{w}$. The Pauli vector can be decomposed in
this system as
\begin{equation} \label{Hermitian decomposition}
\boldsymbol{\sigma}
  =\mathbf{u} (\mathbf{u} \cdot \boldsymbol{\sigma})
  +\mathbf{v} (\mathbf{v} \cdot \boldsymbol{\sigma})
  +\mathbf{w} (\mathbf{w} \cdot \boldsymbol{\sigma}),
\end{equation}
where $\mathbf{u} \cdot \boldsymbol{\sigma}$, $\mathbf{v} \cdot \boldsymbol{\sigma}$, and
$\mathbf{w} \cdot \boldsymbol{\sigma}$ anti-commute with one another and satisfy the
following relations,
\begin{subequations} \label{commutations}
\begin{align}
(\mathbf{u} \cdot \boldsymbol{\sigma})(\mathbf{v} \cdot \boldsymbol{\sigma})
    &= -(\mathbf{v} \cdot \boldsymbol{\sigma})(\mathbf{u} \cdot \boldsymbol{\sigma})
    =i \mathbf{w} \cdot \boldsymbol{\sigma}, \\
(\mathbf{v} \cdot \boldsymbol{\sigma})(\mathbf{w} \cdot \boldsymbol{\sigma})
    &= -(\mathbf{w} \cdot \boldsymbol{\sigma})(\mathbf{v} \cdot \boldsymbol{\sigma})
    =i \mathbf{u} \cdot \boldsymbol{\sigma}, \\
(\mathbf{w} \cdot \boldsymbol{\sigma})(\mathbf{u} \cdot \boldsymbol{\sigma})
    &= -(\mathbf{u} \cdot \boldsymbol{\sigma})(\mathbf{w} \cdot \boldsymbol{\sigma})
    =i \mathbf{v} \cdot \boldsymbol{\sigma}.
\end{align}
\end{subequations}
In addition, orthogonal unit vectors $\mathbf{u}$ and $\mathbf{v}$ form a basis to span
the two-dimensional vector space that is perpendicular to unit vector $\mathbf{w}$. From
this basis we construct a pair of mutually orthogonal complex unit vectors
$\mathbf{w}_{\pm}$ as follows,
\begin{subequations} \label{complex basis}
\begin{align}
\mathbf{w}_{+}(\mathbf{I}) & =\frac{1}{\sqrt{2}}(\mathbf{u}+ i\mathbf{v}), \\
\mathbf{w}_{-}(\mathbf{I}) & =\frac{1}{\sqrt{2}}(\mathbf{v}+ i\mathbf{u}),
\end{align}
\end{subequations}
and define a pair of ladder operators,
\begin{equation} \label{ladder operators}
\sigma^w_{\pm}(\mathbf{I}) =\mathbf{w}_{\pm}(\mathbf{I}) \cdot \boldsymbol{\sigma},
\end{equation}
where the dependence of each quantity on unit vector $\mathbf{I}$ is explicitly
indicated. From Eqs. (\ref{commutations}) and (\ref{complex basis}), one readily write
down the following relations,
\begin{subequations} \label{property 1}
\begin{align}
\sigma^w_{+} (\mathbf{I}) (\mathbf{w} \cdot \boldsymbol{\sigma}) & =
    -\sigma^w_{+} (\mathbf{I}), \\
\sigma^w_{-} (\mathbf{I}) (\mathbf{w} \cdot \boldsymbol{\sigma}) & =
     \sigma^w_{-} (\mathbf{I}).
\end{align}
\end{subequations}
They show that if a spinor $\chi$ is an eigen function of $\mathbf{w} \cdot
\boldsymbol{\sigma}$ with eigen value $+1$, one has $\sigma^w_{+} \chi=0$. Similarly, if
$\chi$ is an eigen function of $\mathbf{w} \cdot \boldsymbol{\sigma}$ with eigen value
$-1$, one has $\sigma^w_{-} \chi=0$. Furthermore, from Eqs. (\ref{commutations}) and
(\ref{complex basis}) one also has
\begin{subequations} \label{property 2}
\begin{align}
(\mathbf{w} \cdot \boldsymbol{\sigma}) \sigma^w_{+} (\mathbf{I}) & =
    \sigma^w_{+} (\mathbf{I}), \\
(\mathbf{w} \cdot \boldsymbol{\sigma}) \sigma^w_{-} (\mathbf{I}) & =
   -\sigma^w_{-} (\mathbf{I}).
\end{align}
\end{subequations}
The first equation indicates that for any spinor $\chi$ that satisfies $\sigma^w_{+} \chi
\neq 0$, $\sigma^w_{+} \chi$ is the eigen function of $\mathbf{w} \cdot
\boldsymbol{\sigma}$ with eigen value $+1$. And the second equation indicates that for
any spinor $\chi$ that satisfies $\sigma^w_{-} \chi \neq 0$, $\sigma^w_{-} \chi$ is the
eigen function of $\mathbf{w} \cdot \boldsymbol{\sigma}$ with eigen value $-1$.
Consequently, the two normalized solutions to equation (\ref{basic spinors}) have the
form of
\begin{subequations} \label{eigen spinors}
\begin{align}
\chi^w_{+}(\mathbf{I}) & = N_{+} \sigma^w_{+}(\mathbf{I})
    \chi_1 \label{eigen spinor 1}, \\
\chi^w_{-}(\mathbf{I}) & = N_{-} \sigma^w_{-}(\mathbf{I})
    \chi_2 \label{eigen spinor 2},
\end{align}
\end{subequations}
where $\chi_1$ and $\chi_2$ are any two fixed spinors satisfying $\sigma^w_{+}
(\mathbf{I}) \chi_1 \neq 0$ and $\sigma^w_{-} (\mathbf{I}) \chi_2 \neq 0$, respectively,
and
\begin{eqnarray}
N_{+} & = & \frac{1}{[\chi^{\dag}_1 (1- \mathbf{w} \cdot \boldsymbol{\sigma})
            \chi_1]^{1/2}}, \nonumber \\
N_{-} & = & \frac{1}{[\chi^{\dag}_2 (1+ \mathbf{w} \cdot \boldsymbol{\sigma})
            \chi_2]^{1/2}}  \nonumber
\end{eqnarray}
are the normalization coefficients which are independent of $\mathbf I$.

\subsection{Unit vector $\mathbf I$ characterizes the phase of eigen spinors $\chi^w_{\pm}$}

\begin{figure}[ht]
\includegraphics{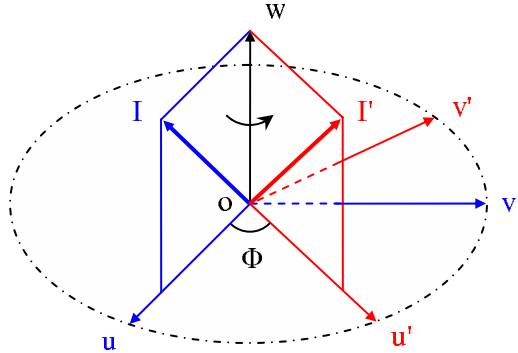}
\caption{(Color online) Schematic diagram for the rotation of unit vector $\mathbf{I}$
about $\mathbf{w}$ axis through an angle $\Phi$.} \label{azimuthal angle}
\end{figure}
Spinors in Eqs. (\ref{eigen spinors}) are always eigen functions no matter how one
chooses the unit vector $\mathbf{I}$. That is to say, they have degrees of freedom
represented by the unit vector $\mathbf{I}$. In the following we will show that this unit
vector plays the role of determining the phases of the eigen spinors. According to Eq.
(\ref{v}), only the azimuthal angle of $\mathbf{I}$ with respect to $\mathbf{w}$ has
effect on vectors $\mathbf{u}$ and $\mathbf{v}$. So the unit-vector degree of freedom in
this case reduces to this azimuthal angle which varies with the rotation of $\mathbf{I}$
about $\mathbf{w}$, as is shown in Fig. \ref{azimuthal angle}. For the sake of
convenience, we write out the matrices of two different rotations. The matrix of a SO(3)
rotation $R(\phi \mathbf{w}) =\exp(-i \phi \mathbf{w} \cdot \mathbf{\Sigma})$ through an
angle $\phi$ about the $\mathbf{w}$ axis can be written as \cite{Normand}
\begin{equation} \label{so3 rotation}
R(\phi \mathbf{w})
  =\cos\phi- i \mathbf{w} \cdot \mathbf{\Sigma} \sin\phi
  +(1-\cos\phi) \mathbf{w} \mathbf{w}^{T},
\end{equation}
where the superscript $T$ means transpose, and the components of matrix vector
$\mathbf{\Sigma}$ are given by
$$
\Sigma_x=\left(\begin{array}{ccc} 0 & 0 & 0  \\
                                  0 & 0 & -i \\
                                  0 & i & 0  \end{array}\right), \hspace{0pt}
\Sigma_y=\left(\begin{array}{ccc} 0  & 0 & i  \\
                                  0  & 0 & 0 \\
                                  -i & 0 & 0  \end{array}\right), \hspace{0pt}
\Sigma_z=\left(\begin{array}{ccc} 0 & -i & 0  \\
                                  i & 0  & 0 \\
                                  0 & 0  & 0  \end{array}\right).
$$
And the matrix of a SU(2) rotation $U(\phi \mathbf{w}) =\exp(-i \frac{\phi}{2} \mathbf{w}
\cdot \boldsymbol{\sigma})$ through an angle $\phi$ about the $\mathbf{w}$ axis can be
written as \cite{Sakurai}
\begin{equation} \label{su2 rotation}
U(\phi \mathbf{w})
  =\cos\frac{\phi}{2}- i \mathbf{w} \cdot \boldsymbol{\sigma} \sin\frac{\phi}{2}.
\end{equation}
They are related to each other via \cite{Normand}
\begin{equation} \label{2-to-1 correspondence}
[R(\phi \mathbf{w}) \mathbf{a}] \cdot \boldsymbol{\sigma}
    =U(\phi\mathbf{w}) (\mathbf{a} \cdot \boldsymbol{\sigma})
     U^{\dag}(\phi\mathbf{w}),
\end{equation}
where $\mathbf{a}$ is an arbitrary 3-component vector. We have the following theorem:

\noindent Theorem 1: When unit vector $\mathbf{I}$ is transformed as
\begin{equation} \label{rotation of I}
\mathbf{I}'= R(\Phi \mathbf{w}) \mathbf{I}
\end{equation}
by a SO(3) rotation through an angle $\Phi$, eigen spinors $\chi^w_{+} (\mathbf I)$ and
$\chi^w_{-} (\mathbf I)$ are transformed as
\begin{subequations} \label{rotation of eigen spinors}
\begin{align}
\chi^w_{+} (\mathbf{I}') & =U(2\Phi \mathbf{w}) \chi^w_{+}
    (\mathbf{I})= e^{-i\Phi} \chi^w_{+}(\mathbf{I}) \label{rotation of eigen spinor 1}, \\
\chi^w_{-} (\mathbf{I}') & =U(2\Phi \mathbf{w}) \chi^w_{-}
    (\mathbf{I})= e^{i\Phi}  \chi^w_{-}(\mathbf{I}) \label{rotation of eigen spinor 2},
\end{align}
\end{subequations}
respectively, by a SU(2) rotation through twice the angle, $2\Phi$.

Let us prove transformation (\ref{rotation of eigen spinor 1}). In the first place, we
have the following formula \cite{Schiff}:
\begin{equation} \label{action of Sigma}
\mathbf{a} \times \mathbf{b}= -i (\mathbf{a} \cdot \mathbf{\Sigma}) \mathbf{b},
\end{equation}
for the cross product of any two (complex) vectors $\mathbf{a}$ and $\mathbf{b}$. It
follows from Eqs. (\ref{action of Sigma}) and (\ref{u and v}) that
\begin{subequations} \label{relation of u and v}
\begin{align}
(\mathbf{w} \cdot \mathbf{\Sigma}) \mathbf{u} & = i \mathbf{v},\\
(\mathbf{w} \cdot \mathbf{\Sigma}) \mathbf{v} & =-i \mathbf{u}.
\end{align}
\end{subequations}
Since $\mathbf{w}_{+} (\mathbf I)$ is transformed in the same way as $\mathbf{I}$ is,
$\mathbf{w}_{+} (\mathbf{I}')= R(\Phi \mathbf{w}) \mathbf{w}_{+} (\mathbf I)$,
$\chi^w_{+}(\mathbf{I})$ is transformed as, according to Eqs. (\ref{eigen spinor 1}),
(\ref{ladder operators}), (\ref{so3 rotation}), and (\ref{relation of u and v}),
$$
\chi^w_{+} (\mathbf{I}')=\exp(-i \Phi) \chi^w_{+}(\mathbf{I}).
$$
In the second place, because $\chi^w_{+}(\mathbf{I})$ is the eigen spinor of $\mathbf{w}
\cdot \boldsymbol{\sigma}$ with eigen value $+1$, it follows that
$$
U(2\Phi \mathbf{w}) \chi^w_{+}(\mathbf{I})= \exp(-i \Phi) \chi^w_{+}(\mathbf{I}).
$$
Transformation (\ref{rotation of eigen spinor 1}) is thus proven. Transformation
(\ref{rotation of eigen spinor 2}) can be proven in the same way.

We see from Theorem 1 that the rotation of $\mathbf I$ results in a change of
$\chi^w_{+}$ and $\chi^w_{-}$ in their phases. This shows that the azimuthal angle of
$\mathbf I$ with respect to $\mathbf w$ determines the phases of $\chi^w_{+}$ and
$\chi^w_{-}$, and the role of the fixed spinors $\chi_1$ and $\chi_2$ in Eqs. (\ref{eigen
spinor 1}) and (\ref{eigen spinor 2}) is to set a phase reference for $\chi^w_{+}$ and
$\chi^w_{-}$, respectively.

\section{\label{description of SPV} Unit vector $\mathbf{I}$ as degree of freedom
to characterize SPV}

As we know, normalized spinors $\chi^w_{+}$ and $\chi^w_{-}$ in Eqs. (\ref{eigen
spinors}) form an orthonormal basis to express an arbitrary spinor,
\begin{equation} \label{any spinor}
\chi^w (\mathbf{I})= \alpha_1 \chi^w_{+}(\mathbf{I})+ \alpha_2 \chi^w_{-}(\mathbf{I})
    \equiv \varpi^{w} (\mathbf{I}) \tilde{\alpha},
\end{equation}
where $\varpi^{w}= (\begin{array}{lr} \chi^w_{+} & \chi^w_{-} \end{array})$ is referred
to as the mapping matrix, and the coefficients $\alpha_1$ and $\alpha_2$ make up a spinor
$\tilde{\alpha}= \left(\begin{array}{c} \alpha_1 \\ \alpha_2 \end{array} \right)$
satisfying the normalization condition $\tilde{\alpha}^{\dag} \tilde{\alpha} =1$. Since
$\tilde \alpha$ plays a similar role as the Jones vector plays in optics, we will call
$\tilde \alpha$ the generalized Jones vector. Now that $\chi^w_{+}$ and $\chi^w_{-}$
depend on a same unit vector $\mathbf I$, spinor (\ref{any spinor}) varies with this
vector when $\tilde{\alpha}$ is fixed. The purpose of this section is to discuss the
physical meaning of $\mathbf I$ in spinor (\ref{any spinor}).

According to Eq. (\ref{property of SPV}), spinor (\ref{any spinor}) always corresponds to
some SPV $\mathbf{s}(\mathbf{I})$ satisfying
\begin{equation} \label{SPV s}
\mathbf{s}(\mathbf{I}) \cdot \boldsymbol{\sigma} \chi^w (\mathbf{I})= \chi^w
(\mathbf{I}).
\end{equation}
We have the following theorem:

\noindent Theorem 2: When unit vector $\mathbf{I}$ is transformed as Eq. (\ref{rotation
of I}) by a SO(3) rotation through an angle $\Phi$, spinor (\ref{any spinor}) is
transformed as
\begin{equation} \label{rotation of spinor}
\chi^w (\mathbf{I}')=U(2 \Phi \mathbf{w}) \chi^w (\mathbf{I})
\end{equation}
by a SU(2) rotation through an angle $2 \Phi$, and the corresponding SPV is transformed
as
\begin{equation} \label{rotation of SPV}
\mathbf{s}(\mathbf{I}')= R(2\Phi \mathbf{w})\mathbf{s}(\mathbf{I})
\end{equation}
by a SO(3) rotation also through an angle $2 \Phi$.

The proof is easily given. First of all, transformation (\ref{rotation of spinor})
follows directly from Theorem 1. In addition, From Eq. (\ref{2-to-1 correspondence}) one
has
\begin{equation} \label{rotation of SPO}
[R(2\Phi\mathbf{w})\mathbf{s}(\mathbf{I})] \cdot \boldsymbol{\sigma}
    =U(2\Phi\mathbf{w}) [\mathbf{s}(\mathbf{I}) \cdot \boldsymbol{\sigma}]
    U^{\dag}(2\Phi\mathbf{w}).
\end{equation}
Acting both sides of this equation on $\chi(\mathbf{I}')$ and taking Eqs. (\ref{SPV s})
and (\ref{rotation of spinor}) into account, one obtains
\begin{equation} \label{eigen equation}
\{ [R(2\Phi\mathbf{w})\mathbf{s}(\mathbf{I})] \cdot \boldsymbol{\sigma} \}
    \chi^w (\mathbf{I}')=\chi^w (\mathbf{I}'),
\end{equation}
indicating that the SPV corresponding to spinor $\chi^w (\mathbf{I}')$ is $R (2\Phi
\mathbf{w}) \mathbf{s} (\mathbf{I})$. This proves Theorem 2.

Eq. (\ref{rotation of spinor}) states that if $\mathbf{I}$ is rotated by an angle, spinor
$\chi^w (\mathbf I)$ is rotated by twice the angle. Therefore, if $\mathbf{I}$ is rotated
by $2\pi$, the spinor recovers to its initial value. Eq. (\ref{rotation of SPV}) states
that if $\mathbf{I}$ is rotated by an angle, the SPV corresponding to the spinor is
rotated by twice the angle, the same as the spinor itself is rotated. This result is
rooted in Eq. (\ref{2-to-1 correspondence}), a direct consequence of the equivalence
between the SO(3) and SU(2) rotations as will be discussed below. It deserves
reemphasizing that the polar angle of $\mathbf I$ with respect to $\mathbf w$ is a
degenerate degree of freedom in this case. Only the azimuthal angle plays the role.

It is noted that the mapping matrix $\varpi^w (\mathbf I)$ introduced in Eq. (\ref{any
spinor}) is a unitary matrix. It transforms the generalized Jones vector $\tilde
{\alpha}$, a fixed spinor, into an $\mathbf I$-dependent spinor $\chi^w$. This is the
point of view in Schr\"{o}dinger picture. In the next, we will explain the transformation
(\ref{rotation of SPV}) with the language of Heisenberg picture. When working in
Heisenberg picture, the fixed Pauli vector (\ref{Hermitian decomposition}), a dynamical
variable, is transformed by $\varpi^w (\mathbf I)$ into an $\mathbf I$-dependent Pauli
vector,
\begin{equation} \label{Pauli vector in HP}
\boldsymbol{\sigma}^{H}(\mathbf I)
  =\varpi^{w \dag} (\mathbf I) \boldsymbol{\sigma} \varpi^w (\mathbf{I})
  =\mathbf{u} \sigma^{H}_{u}(\mathbf I)
  +\mathbf{v} \sigma^{H}_{v}(\mathbf I)
  +\mathbf{w} \sigma^{H}_{w}(\mathbf I),
\end{equation}
where
\begin{subequations} \label{definition of Pauli matrix}
\begin{align}
\sigma^{H}_{u}(\mathbf I)
  &=\varpi^{w \dag}(\mathbf I) (\mathbf{u} \cdot \boldsymbol{\sigma}) \varpi^w (\mathbf{I}),\\
\sigma^{H}_{v}(\mathbf I)
  &=\varpi^{w \dag}(\mathbf I) (\mathbf{v} \cdot \boldsymbol{\sigma}) \varpi^w (\mathbf{I}),\\
\sigma^{H}_{w}(\mathbf I)
  &=\varpi^{w \dag}(\mathbf I) (\mathbf{w} \cdot \boldsymbol{\sigma}) \varpi^w (\mathbf{I}).
\end{align}
\end{subequations}
From Eq. (\ref{complex basis}), one finds
\begin{subequations} \label{unitary transform}
\begin{align}
\mathbf{u} \cdot \boldsymbol{\sigma} & = \frac{1}{\sqrt 2} (\sigma^w_{+} -i\sigma^w_{-}), \\
\mathbf{v} \cdot \boldsymbol{\sigma} & =-\frac{i}{\sqrt 2} (\sigma^w_{+} +i\sigma^w_{-}).
\end{align}
\end{subequations}
In addition, according to Eq. (\ref{property 2}), one has
\begin{subequations} \label{ladder operations}
\begin{align}
\sigma^w_{+} \chi^w_{-} & =c  \chi^w_{+}, \\
\sigma^w_{-} \chi^w_{+} & =c' \chi^w_{-},
\end{align}
\end{subequations}
which lead to
\begin{equation} \label{c}
c=i c'^* =2N_{+}N_{-} \chi^{\dag}_1 \sigma^w_{-} \chi_2,
\end{equation}
where $|c|=|c'|=\sqrt 2$. Denoting $c= \sqrt{2} \exp(i \varphi_0)$, one obtains
\begin{equation} \label{reference phase factor}
\exp(i \varphi_0)= \sqrt{2} N_{+}N_{-} \chi^{\dag}_1 \sigma^w_{-} \chi_2,
\end{equation}
which depends on unit vector $\mathbf I$ through operator $\sigma^w_{-} (\mathbf I)$.
Substituting $\varpi^{w}= (\begin{array}{lr} \chi^w_{+} & \chi^w_{-}
\end{array})$ into Eq. (\ref{definition of Pauli matrix}) and taking Eqs. (\ref{unitary
transform})-(\ref{reference phase factor}) into account, we finally find
\begin{subequations} \label{Pauli matrix}
\begin{align}
\sigma^{H}_{u}(\mathbf I)
  &=\left(\begin{array}{cc} 0 &   e^{i\varphi_0}\\ e^{-i\varphi_0} & 0 \end{array}\right),\\
\sigma^{H}_{v}(\mathbf I)
  &=\left(\begin{array}{cc} 0 & -ie^{i\varphi_0}\\ie^{-i\varphi_0} & 0 \end{array} \right), \\
\sigma^{H}_{w}(\mathbf I)
  & =\left(\begin{array}{cc} 1 & 0 \\ 0 & -1 \end{array} \right).
\end{align}
\end{subequations}

Eq. (\ref{Pauli vector in HP}) together with Eq. (\ref{Pauli matrix}) represents an
initial value of the Pauli vector. After $\mathbf I$ is changed into $\mathbf I'$
according to Eq. (\ref{rotation of I}), the Pauli vector becomes
$$
\boldsymbol{\sigma}^{H}(\mathbf I')
  =\mathbf{u}' \sigma^{H}_{u}(\mathbf I')
  +\mathbf{v}' \sigma^{H}_{v}(\mathbf I')
  +\mathbf{w}' \sigma^{H}_{w}(\mathbf I').
$$
On one hand, it is easy to show that
$$
\exp[i \varphi_0(\mathbf I')]=\exp(i \Phi) \exp[i \varphi_0(\mathbf I)].
$$
Therefore, one has from Eq. (\ref{Pauli matrix})
\begin{subequations} \label{rotated sigma}
\begin{align}
\sigma^{H}_{u}(\mathbf I')
  & =U^{\dag} (\Phi \mathbf w)\sigma^{H}_{u}(\mathbf I) U (\Phi \mathbf w), \\
\sigma^{H}_{v}(\mathbf I')
  & =U^{\dag} (\Phi \mathbf w)\sigma^{H}_{v}(\mathbf I) U (\Phi \mathbf w), \\
\sigma^{H}_{w}(\mathbf I')
  & =U^{\dag} (\Phi \mathbf w)\sigma^{H}_{w}(\mathbf I) U (\Phi \mathbf w)
    =\sigma^{H}_{w}(\mathbf I).
\end{align}
\end{subequations}
On the other hand, unit vectors $\mathbf u$, $\mathbf v$, and $\mathbf w$ are rotated in
the same way as $\mathbf I$ is rotated. Collecting all these relations, we get
\begin{subequations}
\begin{align}
\boldsymbol{\sigma}^{H}(\mathbf I')
  & =R(\Phi \mathbf w) [U^{\dag} (\Phi \mathbf w) \boldsymbol{\sigma}^{H}
     (\mathbf I) U (\Phi \mathbf w)] \label{rotated PV1}  \\
  & =U^{\dag} (\Phi \mathbf w) [R(\Phi \mathbf w) \boldsymbol{\sigma}^{H}
     (\mathbf I)] U (\Phi \mathbf w),\label{rotated PV2}
\end{align}
\end{subequations}
where $R(\Phi \mathbf w) \boldsymbol{\sigma}^{H}(\mathbf I)$ is meant by
\begin{equation} \label{PV is ordinary vector}
R(\Phi \mathbf w) \boldsymbol{\sigma}^{H}(\mathbf I)
  =[R(\Phi \mathbf w) \mathbf{u}] \sigma^{H}_{u}(\mathbf I)
  +[R(\Phi \mathbf w) \mathbf{v}] \sigma^{H}_{v}(\mathbf I)
  +[R(\Phi \mathbf w) \mathbf{w}] \sigma^{H}_{w}(\mathbf I).
\end{equation}
Eq. (\ref{PV is ordinary vector}) shows that the Pauli vector transforms as an ordinary
vector under the SO(3) rotation. Furthermore, Eq. (\ref{rotation of spinor}) means
$\chi^w (\mathbf I')=\varpi^w (\mathbf I') \tilde {\alpha}$, where $\varpi^w (\mathbf I')
=U(2 \Phi \mathbf{w}) \varpi^w (\mathbf I)$. It follows from Eq. (\ref{Pauli vector in
HP}) that the Pauli vector that corresponds to $\mathbf I'$ can also be written as
\begin{equation} \label{su rotation of PV}
\boldsymbol {\sigma}^{H} (\mathbf I')
  =U^{\dag} (2\Phi \mathbf w) \boldsymbol{\sigma}^{H}(\mathbf I) U (2\Phi \mathbf w).
\end{equation}
Comparing Eq. (\ref{su rotation of PV}) with Eq. (\ref{rotated PV2}), one finds
\begin{equation} \label{equivalence}
R(\Phi \mathbf w) \boldsymbol{\sigma}^{H}(\mathbf I)
  =U^{\dag} (\Phi \mathbf w) \boldsymbol{\sigma}^{H}(\mathbf I) U (\Phi \mathbf w).
\end{equation}
This is the generalization of Eq. (\ref{2-to-1 correspondence}), conveying an equivalence
\cite{Damask} between the SO(3) and SU(2) rotations when operating onto the Pauli vector.
Indeed, it is easy to derive Eq. (\ref{2-to-1 correspondence}) from Eq.
(\ref{equivalence}). With an arbitrary vector $\mathbf a$, one has from Eq.
(\ref{equivalence})
$$
[R(\Phi \mathbf w) R^{\dag}(\Phi \mathbf w) \mathbf{a}] \cdot [R(\Phi \mathbf w)
   \boldsymbol{\sigma}^{H}(\mathbf I)]
  =U^{\dag} (\Phi \mathbf w) [\mathbf{a} \cdot \boldsymbol{\sigma}^{H}(\mathbf I)]
   U(\Phi \mathbf w),
$$
which reduces to
$$
[R^{\dag}(\Phi \mathbf w) \mathbf{a}] \cdot \boldsymbol{\sigma}^{H}(\mathbf I)
  =U^{\dag} (\Phi \mathbf w) [\mathbf{a} \cdot \boldsymbol{\sigma}^{H}(\mathbf I)]
   U(\Phi \mathbf w),
$$
due to the fact that a rotation does not change the scalar product of any two vectors.
Once making substitution $\Phi =-\phi$ and noticing properties $R(-\phi \mathbf w)
=R^{\dag}(\phi \mathbf w)$ and $U(-\phi \mathbf w) =U^{\dag}(\phi \mathbf w)$, one
obtains
$$
[R(\phi \mathbf{w}) \mathbf{a}] \cdot \boldsymbol{\sigma}^{H}(\mathbf I)
    =U(\phi\mathbf{w}) [\mathbf{a} \cdot \boldsymbol{\sigma}^{H}(\mathbf I)]
     U^{\dag}(\phi\mathbf{w}).
$$
This is nothing but equation (\ref{2-to-1 correspondence}). Such an equivalence allows us
to rewrite Eq. (\ref{su rotation of PV}) as
\begin{equation} \label{rotation of PV}
\boldsymbol {\sigma}^{H} (\mathbf I')
  =R(2\Phi \mathbf{w}) \boldsymbol {\sigma}^{H} (\mathbf I),
\end{equation}
which can also be obtained by substituting Eq. (\ref{equivalence}) into Eq. (\ref{rotated
PV1}).

Eq. (\ref{rotation of PV}) shows that the transformation of the Pauli vector can be
expressed with a SO(3) rotation through an angle $2 \Phi$ when $\mathbf I$ is rotated
through an angle $\Phi$. Transformation (\ref{rotation of SPV}) of the SPV follows
directly from this equation.

\section{\label{application} Representation of the spin polarization of free electron}

Now we are ready to formulate a representation of free electron's spin polarization. In
order to explicitly incorporate the spin property, we write the non-relativistic
Hamiltonian of a free electron as
\begin{equation} \label{Hamiltonian}
H= \frac{(\mathbf{p} \cdot \boldsymbol{\sigma})^2}{2 \mu}.
\end{equation}
Since the momentum $\mathbf{p}$ commutes with $H$, a spinor solution to the
Schr\"{o}dinger equation can be expressed as an integral over a plane-wave spectrum,
\begin{equation} \label{beam}
\Psi(\mathbf{x},t; \mathbf{I})= \frac{1}{(2\pi)^{3/2}}
    \int A(\mathbf k) \chi^w (\mathbf{I})
    e^{i (\mathbf{k} \cdot \mathbf{x}-\omega t)} d^3 k,
\end{equation}
where the quantization axis for each plane wave is assumed to be the wave vector
\cite{Beenakker}, $\mathbf{w}= \mathbf{k}/ |\mathbf{k}|$, $\chi^w (\mathbf{I})$ is given
by Eq. (\ref{any spinor}), $\omega =\frac{\hbar^2 k^2} {2\mu}$, and $A(\mathbf k)$ is the
weighting function satisfying the normalization condition $\int |A(\mathbf k)|^2 d^3
k=1$. The unit vector $\mathbf I$ in Eq. (\ref{beam}) is assumed to be the same to all
the plane wave and thus functions as an independent index to characterize the spinor wave
function $\Psi(\mathbf{x},t; \mathbf{I})$. Substituting Eq. (\ref{any spinor}) into Eq.
(\ref{beam}), one has
\begin{equation} \label{any beam}
\Psi(\mathbf{x},t;\mathbf{I}) =\Pi^p (\mathbf{x},t;\mathbf{I}) \tilde{\alpha},
\end{equation}
where $\Pi^p =(\begin{array}{lr} \Psi^p_{+} & \Psi^p_{-} \end{array})$, and
\begin{subequations} \label{eigen functions}
\begin{align}
\Psi^p_{+}(\mathbf{x},t;\mathbf{I})
  & = \frac{1}{(2\pi)^{3/2}} \int A(\mathbf k) \chi^w_{+} (\mathbf{I})
     e^{i(\mathbf{k} \cdot \mathbf{x}-\omega t)} d^3 k, \label{eigen function 1} \\
\Psi^p_{-}(\mathbf{x},t;\mathbf{I})
  & = \frac{1}{(2\pi)^{3/2}} \int A(\mathbf k) \chi^w_{-} (\mathbf{I})
     e^{i(\mathbf{k} \cdot \mathbf{x}-\omega t)} d^3 k.
\end{align}
\end{subequations}
Here $\Psi^p_{+}$ and $\Psi^p_{-}$ are characterized by the same unit vector $\mathbf I$.

It is easy to prove that $\Psi^p_{+}$ and $\Psi^p_{-}$ are eigen functions of operator
$\mathbf{w}^p \cdot \boldsymbol{\sigma}$,
\begin{equation}\label{eigen spinors of SMC}
\mathbf{w}^p \cdot \boldsymbol{\sigma} \Psi^p_{\pm} =\pm \Psi^p_{\pm},
\end{equation}
having eigen values $+1$ and $-1$, respectively, where $\mathbf{w}^p =\frac{\mathbf
p}{p}$ is the unit momentum operator. But now different $\mathbf I$'s will correspond
different eigen functions. Let us consider $\Psi^p_{+}(\mathbf{x},t;\mathbf{I})$ as an
example. According to Theorem 1, unit vector $\mathbf I$ determines the phase of spinor
$\chi^w_{+} (\mathbf I)$. Noticing that the weighting function $A(\mathbf k)$ contains in
general different wave vectors in direction, unit vector $\mathbf I$ behaves as a
joystick to control all the spinors $\chi^w_{+} (\mathbf I)$. Changing $\mathbf I$ will
alter their phases differently in accordance with Eq. (\ref{eigen spinor 1}) and
therefore lead to a dissimilar eigen function due to the integral over the wave vector in
Eq. (\ref{eigen function 1}). This is also true of eigen function
$\Psi^p_{-}(\mathbf{x},t;\mathbf{I})$.

According to definition (\ref{definition of SPV}), the local SPV in position space is
represented by spinor $\psi(\mathbf{x},t;\mathbf{I})= \frac{1}{\sqrt{\rho}} \Psi
(\mathbf{x},t; \mathbf{I})$, which satisfies $\psi^{\dag} \psi=1$, where $\rho=
\Psi^{\dag} \Psi$ is the probability density. Correspondingly, the local SPV $\mathbf{s}
(\mathbf{x},t;\mathbf{I}) =\psi^{\dag} \boldsymbol{\sigma} \psi$ is given by
\begin{equation}\label{local SPV}
\mathbf{s} (\mathbf{x},t;\mathbf{I})
  =\frac{1}{\rho} \Psi^{\dag}(\mathbf{x},t;\mathbf{I}) \boldsymbol{\sigma}
   \Psi(\mathbf{x},t;\mathbf{I}).
\end{equation}
Since the spin angular momentum is $\frac{\hbar}{2}$ times the Pauli vector, $\frac
{\hbar} {2} \rho \mathbf{s} (\mathbf{x},t;\mathbf{I})$ is the density of spin angular
momentum. By making use of Eq. (\ref{any beam}), we have for the total spin angular
momentum,
\begin{equation}
\mathbf{S} (\mathbf I)
  =\frac{\hbar}{2} \int \rho \mathbf{s} (\mathbf{x},t;\mathbf{I}) d^3 x
  =\frac{\hbar}{2} \tilde{\alpha}^{\dag} \int \Pi^{p \dag} \boldsymbol{\sigma}
   \Pi^p d^3 x \tilde{\alpha}.
\end{equation}
With the help of Eq. (\ref{eigen functions}), it becomes
\begin{equation}\label{total spin}
\mathbf{S} (\mathbf I)
  =\frac{\hbar}{2} \tilde{\alpha}^{\dag} \int |A(\mathbf k)|^2 \boldsymbol{\sigma}^H
   (\mathbf{I}) d^3 k \tilde{\alpha}.
\end{equation}
which is independent of the time, where $\boldsymbol{\sigma}^H (\mathbf{I})$ is given by
(\ref{Pauli vector in HP}). Eq. (\ref{local SPV}) shows that the local SPV is dependent
on $\mathbf I$. Eq (\ref{total spin}) shows that the total spin is in general dependent
on $\mathbf I$, too.

We summarize our representation of free electron's spin polarization as follows. The two
eigen functions of $\mathbf{w}^p \cdot \boldsymbol{\sigma}$ form the basis of this
representation. They are characterized by the same weighting function $A(\mathbf k)$ and
the same unit vector $\mathbf I$. Each state of spin polarization in this representation
is expressed by wave function (\ref{any beam}). It is described by three kinds of
parameters, the generalized Jones vector $\tilde{\alpha}$, the weighting function
$A(\mathbf k)$, and the unit vector $\mathbf I$. They are independent of one another.
$\tilde{\alpha}$ and $A(\mathbf k)$ have the traditional meaning. But the unit vector
$\mathbf I$ is identified here for the first time. When $\tilde{\alpha}$ and $A(\mathbf
k)$ are given, the spin polarization is completely determined by $\mathbf I$.

\section{\label{conclusions} Conclusions and remarks}

In conclusion, we showed that the azimuthal angle of the unit vector $\mathbf I$ with
respect to the quantization direction plays the role of characterizing the spin
polarization if only the spin of electron is considered. This result is mainly expressed
by Eqs. (\ref{rotation of I}) and (\ref{rotation of spinor}). Upon utilizing this
approach to a free electron, we formulate a representation which consists of a pair of
polarization bases and is characterized by a unit vector. When the weighting function and
generalized Jones vector are given, the local SPV and the total spin angular momentum are
completely determined by this vector.

Equations (\ref{rotation of spinor}) and (\ref{rotation of I}) may suggest an isomorphism
between the SU(2) and SO(3) rotation groups, a one-to-one correspondence that is
different from the well known two-to-one covering relation in the literature
\cite{Normand, Sakurai, Schiff}. In this correspondence, the SU(2) rotation operates on
2-component spinors; the SO(3) rotation operates on 3-component vectors. But the angle of
the SU(2) rotation is twice the angle of the SO(3) rotation.

Our representation was formulated within the framework of non-relativistic quantum
mechanics. It is not difficult to generalize to the case of relativistic quantum
mechanics upon replacing the Pauli vector with the relativistic spin operators
\cite{Rose}.

\section*{Acknowledgments}

The authors are indebted to Xi Chen for helpful suggestions. This work was supported in
part by the National Natural Science Foundation of China (60877055 and 60806041), the
Science and Technology Commission of Shanghai Municipal (08JC14097 and 08QA14030), the
Shanghai Educational Development Foundation (2007CG52), and the Shanghai Leading Academic
Discipline Project (S30105).

\end{document}